\documentclass[12pt]{article}
\usepackage{amssymb,amsmath,epsfig}
\allowdisplaybreaks
\begin{document}
\title{\bf Impact of Kuchowicz Metric Function on Gravastars in $f(R,T)$ Theory}

\author{M. Sharif \thanks{msharif.math@pu.edu.pk} and Arfa Waseem
\thanks{arfawaseem.pu@gmail.com}\\
Department of Mathematics, University of the Punjab,\\
Quaid-e-Azam Campus, Lahore-54590, Pakistan.}
\date{}

\maketitle

\begin{abstract}
This paper discusses the configuration of gravitational vacuum star
or gravastar with the impact of geometry and matter coupling present
in $f(R,T)$ gravity. The gravastar is also conceptualized as a
substitute for a black hole which is illustrated by three geometries
known as (1) the interior geometry, (2) the intermediate thin-shell
and (3) the exterior geometry. For a particular $f(R,T)$ model, we
analyze these geometries corresponding to Kuchowicz metric function.
We evaluate another metric potential for the interior domain as well
as the intermediate shell which is non-singular for both domains.
The Schwarzschild metric is adopted to demonstrate the exterior
geometry of gravastar, while the numerical values of unknown
constants are calculated through boundary conditions. Finally, we
discuss different features of gravastar regions like proper length,
energy, surface redshift as well as equation of state parameter. We
conclude that the gravastar model can be regarded as a successful
replacement of the black hole in the context of this gravity.
\end{abstract}

\section{Introduction}

There exist enormous astrophysical phenomena including gravitational
collapse as well as stellar evolution that have become the center of
attention for many researchers. Stellar evolution describes a
process that manifests the entire life cycle of a self-gravitating
object whereas the phenomenon of gravitational collapse is the cause
behind the production of new and massive compact celestial bodies
like neutron stars, white dwarfs and black holes. The last result of
this collapsing phenomenon is entirely based upon the parent masses
of stellar objects. Recently, several researchers have proposed that
densest objects other than black holes can be created as the outcome
of massive collapsing stars. In this regard, Mazur and Mottola
\cite{1} developed an elegant model of an extremely compact object
named as gravitational vacuum star or in short gravastar. Gravastar
is identified as the non-singular extremely compact spherically
symmetric candidate that may be considered as compact as the
classical black hole.

According to this model, the geometry of gravastar is demonstrated
by three different regions where the interior geometry depends upon
the de Sitter core with an equation of state (EoS) $\rho=-p>0$,
surrounded by a thin-shell with stiff matter and the external vacuum
domain is expressed by the Schwarzschild spacetime. The notion of
gravastar geometries is significantly attractive for the
astrophysicists as it may resolve the fundamental issues associated
with black holes, i.e., the information paradox and the issue of
singularity. It is believed that a phase transition occurs in the
interior of gravastar that generates a de Sitter core with repulsive
effects which assist to balance the collapsing body and prevent the
generation of singularity \cite{2}. This transition appears very
close to the bound $\frac{2m}{r}=1$ which makes difficult for an
observer to differentiate the gravastar from a black hole.

Despite several theoretical and observational achievements, there
are still complicated issues that motivate the researchers to
explore different substitutes in which the collapsing results are
very large stars without event horizons \cite{2}-\cite{4}. Among
those giant stars, gravastars have inspired the astrophysicists to
explore their geometries using various techniques. In general
relativity (GR), Visser and Wiltshire \cite{5} examined the
stability of gravastars and observed that different EoS provide the
dynamically stable distribution of gravastars. Carter \cite{6}
established new exact solutions of gravastars and found different
effects of EoS on the exterior and interior regions. Bili\'{c} et
al. \cite{7} investigated the core solutions of gravastars by
inducing the Born-Infeld phantom in place of de Sitter metric and
examined that at the center of galaxies, their results can
illustrate the dark compact structures.

Horvat and Iliji\'{c} \cite {8} analyzed the behavior of energy
bounds within the gravastar shell and discussed the stability
through radial oscillations and sound speed on the shell. Many
researchers \cite{9}-\cite{14} presented exact interior solutions of
gravastar by employing various EoS corresponding to different
conjectures. Lobo and Arellano \cite{15} formulated distinct models
of gravastars under the impact of nonlinear electrodynamics and
inspected some particular characteristics of their models. Horvat et
al. \cite{16} generalized the notion of gravastar by inserting the
effects of electric field and derived the surface redshift, EoS
parameter as well as stability for the inner and outer geometries.
In the same scenario, Turimov et al. \cite{17} examined the magnetic
field configuration and obtained exact solutions corresponding to
the interior geometry of a slowly rotating gravastar.

In cosmological scenarios, the mysteries of dark constituents,
responsible for the expanding cosmos, lead to the formation of
extended or modified gravitational theories to GR. The concept of
coupling between matter and curvature yields several modified
theories like $f(R,T)$ gravity \cite{18}, $f(R,T,
R_{\gamma\eta}T^{\gamma\eta})$ theory \cite{19} and
$f(\mathcal{G},T)$ gravity \cite{20}, where $R$ indicates the
curvature invariant, $T$ reveals the trace of energy-momentum tensor
(EMT) and $\mathcal{G}$ symbolizes the Gauss-Bonnet invariant. In
these theories, $f(R,T)$ theory has attained much importance in
describing various astrophysical objects corresponding to different
approaches \cite{21}-\cite{28}. The study of gravastar has motivated
the researchers to discuss the effects of modified theories on
structural properties of gravastar. In $f(R,T)$ framework, Das et
al. \cite{29} analyzed the speculation of gravastar and studied its
features graphically with respect to different EoS. Shamir and Ahmad
\cite{30} exhibited singularity free solutions of gravastar and
extracted mathematical forms of several physical quantities in
$f(\mathcal{G},T)$ background.

The physical viability as well as stability of gravastar models have
also been analyzed with the impact of anisotropic pressure
configuration. Cattoen et al. \cite{31} expressed the extended form
of gravastar that comprises anisotropic pressure component beyond
the presence of thin-shell. They observed that pressure anisotropy
is the most fundamental ingredient for all gravastar like objects.
DeBenedictis et al. \cite{32} obtained anisotropic interior
solutions of gravastar by considering different density functions
and EoS. Chan et al. \cite{33} analyzed energy conditions of
gravastars corresponding to anisotropic dark energy as well as
phantom energy fluid distribution. They found that
isotropic/anisotropic interior matter configuration can perturb the
creation of gravastar and the production is feasible when the radial
pressure is less than the tangential pressure.

In the analysis of stellar compact structures, the solutions of the
field equations play a significant role. Ghosh et al. \cite{34}
presented the structure of gravastar by considering the Kuchowicz
metric function in GR. They displayed the graphical behavior of
various features and found stable structure of gravastar under the
influence of their considered singularity free metric function. We
have inspected the effects of electric field on the structure of
gravastar in $f(R,T)$ framework by adopting the conformal Killing
vectors \cite{35}. Ghosh with his collaborators \cite{36} determined
a unique solution of gravastar through embedding class 1 technique
by implementing the Karmarkar condition. They obtained exact
solutions of the intermediate shell and evaluated non-singular
finite solutions for the inner domain of gravastar.

In this paper, we evaluate singularity free solutions of gravastar
admitting the Kuchowicz metric function in $f(R,T)$ gravity. We
determine another metric function for the intermediate shell as well
as interior geometry of gravastar. We present the graphical
variations of different features of gravastar shell for specific
$f(R,T)$ model. We adopt the following pattern for the presentation
of this paper. Next section displays the fundamental formulation of
this theory. Section \textbf{3} expresses three geometries of
gravastar with Kuchowicz metric function. The graphical description
of obtained solutions corresponding to the calculated values of
unknown constants and model parameter is presented in section
\textbf{4}. The last section provides the concluding remarks.

\section{Fundamentals of $f(R,T)$ Theory}

The modified action for $f(R,T)$ gravity coupled with Lagrangian
matter density ($L_{M}$) is characterized by \cite{18}
\begin{equation}\label{1}
\mathcal{A}_{f(R,T)}=\int\sqrt{-g} \left[\frac{f(R,
T)}{2\kappa}+L_{M}\right]d^{4}x,
\end{equation}
in which $g$ exhibits the determinant of $g_{\gamma\eta}$ and
$\kappa=\frac{8\pi G}{c^{4}}$ denotes the coupling constant with
speed of light ($c=1$) and gravitational constant ($G=1$). The
$f(R,T)$ field equations are
\begin{eqnarray}\nonumber
f_{R}(R,T)R_{\gamma\eta}&-&\frac{g_{\gamma\eta}f(R,T)}{2}-
(\nabla_{\gamma}\nabla_{\eta}-g_{\gamma\eta}\Box)f_{R}(R,T)
\\\label{2}&=&8\pi T_{\gamma\eta}-(T_{\gamma\eta}
+\Theta_{\gamma\eta})f_{T}(R,T),
\end{eqnarray}
with $\nabla_{\gamma}$ symbolizes the covariant differentiation,
$f_{R}(R,T)=[f(R,T)]_{,R}$,\\$f_{T}(R,T)= [f(R,T)]_{,T}$,
$\Box=g^{\gamma\eta}\nabla_{\gamma}\nabla_{\eta}$ and
$\Theta_{\gamma\eta}$ is specified by
\begin{equation}\label{3}
\Theta_{\gamma\eta}=g^{\beta\nu}\frac{\delta T_{\beta\nu}}{\delta
g^{\gamma\eta}}= g_{\gamma\eta}L_{m}-2T_{\gamma\eta}-
2g^{\beta\nu}\frac{\partial^{2}L_{m}}{\partial
g^{\gamma\eta}\partial g^{\beta\nu}}.
\end{equation}
The covariant derivative of Eq.(\ref{2}) yields
\begin{equation}\label{4}
\nabla^{\gamma}T_{\gamma\eta}=\frac{f_{T}}{8\pi-f_{T}}
\left[(T_{\gamma\eta}+\Theta_{\gamma\eta})\nabla^{\gamma}(\ln
f_{T})-\frac{g_{\gamma\eta}}{2}\nabla^{\gamma}T+\nabla^{\gamma}
\Theta_{\gamma\eta}\right].
\end{equation}
This expression demonstrates that the conservation law in
curvature-matter coupled theories is not executed by the EMT.

To explore the astrophysical structures, the configuration of matter
shows a significant effect. The non-zero elements of EMT manifest
the dynamical constituents possessing distinct physical
significance. To study the features of gravastars, we assume perfect
matter configuration as
\begin{equation}\label{5}
T_{\gamma\eta}=(p+\rho)\mathcal{U}_{\gamma}\mathcal{U}_{\eta}
-pg_{\gamma\eta}.
\end{equation}
Here, $\rho$ symbolizes the density, $p$ indicates the pressure
while $\mathcal{U}_{\gamma}$ reveals the four velocity. For matter
configuration, there exist several choices of $L_{m}$. Here, we
choose $L_{m}=-p$ giving $\frac{\partial^{2}L_{m}}{\partial
g^{\gamma\eta}\partial g^{\mu\upsilon}}=0$ and
$\Theta_{\gamma\eta}=-2T_{\gamma\eta}-pg_{\gamma\eta}$ \cite{18}. We
consider a linear $f(R,T)$ model in which matter and geometric
components are minimally coupled as
\begin{equation}\label{6}
f(R,T)=R+2h(T),
\end{equation}
where $h(T)$ indicates the arbitrary function of the trace of EMT.
In order to obtain the EMT for modified gravitational theory, we
suppose $h(T)=\alpha T$ where $\alpha$ acts as a coupling constant
and $T=\rho-3p$. The corresponding linear form of $f(R,T)$ becomes
\begin{equation}\label{7}
f(R,T)=R+2\alpha T.
\end{equation}

Here, the induction of $T$ yields the more extended form of GR as
compared to $f(R)$ theory. This form was firstly suggested by Harko
et al. \cite{18} and has widely been implemented to explore the
structural properties of self-gravitating bodies. Das et al.
\cite{29} adopted this functional form to evaluate exact singularity
free solutions of collapsing body and explored various physically
viable properties of gravastar. We have used this model to discuss
the features of charged gravastars with conformal Killing vectors
\cite{35}. Insertion of this form in Eq.(\ref{2}) provides
\begin{equation}\label{8}
G_{\gamma\eta}=8\pi T_{\gamma\eta}+\alpha Tg_{\gamma\eta}
+2\alpha(T_{\gamma\eta}+pg_{\gamma\eta}),
\end{equation}
where $G_{\gamma\eta}$ exhibits the Einstein tensor. Equation
(\ref{4}) for $R+2\alpha T$ model takes the form
\begin{equation}\label{9}
\nabla^{\gamma}T_{\gamma\eta}=\frac{-\alpha}{2(4\pi+\alpha)}
\big[g_{\gamma\eta}\nabla^{\gamma}T+2\nabla^{\gamma}
(pg_{\gamma\eta})\big].
\end{equation}
For $\alpha=0$, the usual conserved form of GR can be regained from
this equation.

\section{Exact Solutions in $R+2\alpha T$ Model}

The static spherically symmetric line element for the interior
region is
\begin{equation}\label{10}
ds^{2}_{-}=e^{\chi(r)}dt^{2}-e^{\xi(r)}dr^{2}
-r^{2}d\theta^{2}-r^{2}\sin^{2}\theta d\phi^{2}.
\end{equation}
The $f(R,T)$ field equations yield
\begin{eqnarray}\label{11}
\frac{1}{r^{2}}-e^{-\xi}\left(\frac{1}{r^{2}}-\frac{\xi'}
{r}\right)&=&8\pi\rho+\alpha(3\rho-p),
\\\label{12}
e^{-\xi}\left(\frac{1}{r^{2}}+\frac{\chi'}{r}\right)-\frac{1}
{r^{2}}&=&8\pi p-\alpha(\rho-3p),
\\\label{13}
e^{-\xi}\left(\frac{\chi''}{2}-\frac{\xi'}{2r}+\frac{\chi'}{2r}
+\frac{\chi'^{2}}{4}-\frac{\chi'\xi'}{4}\right)&=&8\pi
p-\alpha(\rho-3p),
\end{eqnarray}
where prime presents differentiation associated with radial
coordinate. We consider the metric function $e^{\chi(r)}$ as
Kuchowicz type \cite{37} expressed by
\begin{equation}\label{14}
e^{\chi(r)}=e^{Ar^{2}+2\ln B},
\end{equation}
where $A$ and $B$ are arbitrary constants. The dimension of $A$ is
$\frac{1}{L^{2}}$ while $B$ is a dimensionless constant. This metric
function was proposed by Kuchowicz as a non-singular function to
discuss the stellar configurations. Ghosh et al. \cite{34} employed
this metric potential to analyze the features of gravastars in GR.
This function provides successful theoretical assistance for massive
compact objects corresponding to some particular values of arbitrary
constants. Substituting Eq.(\ref{14}) in (\ref{11})-(\ref{13}), we
have
\begin{eqnarray}\label{15}
\frac{1}{r^{2}}-e^{-\xi}\left(\frac{1}{r^{2}}-\frac{\xi'}
{r}\right)&=&8\pi\rho+\alpha(3\rho-p),
\\\label{16}
e^{-\xi}\left(\frac{1}{r^{2}}+2A\right)-\frac{1} {r^{2}}&=&8\pi
p-\alpha(\rho-3p),
\\\label{17}
e^{-\xi}\left(2A-\frac{\xi'}{2r}
+A^{2}r^{2}-\frac{Ar\xi'}{2}\right)&=&8\pi p-\alpha(\rho-3p).
\end{eqnarray}
The modified conservation equation turns out to be
\begin{equation}\label{18}
p'+Ar(p+\rho)-\frac{\alpha}{2(\alpha+4\pi)}\Big(\rho'-p'\Big)=0.
\end{equation}

\subsection{Structure of Gravastar}

Here, we explore the geometrical composition of gravastar with
Kuchowicz metric function in $f(R,T)$ scenario. We inspect the
regions of gravastar, i.e., (1) the inner, (2) the intermediate
thin-shell and (3) the outer. The internal domain of gravastar is
enclosed by thin-shell which consists of stiff matter while the
outer domain is entirely a vacuum and the Schwarzschild spacetime
can be identified suitable for this external system. The shell is
assumed to be extremely thin having a finite width satisfying the
range
$\mathcal{R}_{1}=\mathcal{R}<r<\mathcal{R}+\epsilon=\mathcal{R}_{2}$,
in which $\mathcal{R}_{1}$ and $\mathcal{R}_{2}$ indicate the
interior and exterior radii of gravastar, respectively and thickness
of the shell is exhibited by
$\mathcal{R}_{2}-\mathcal{R}_{1}=\epsilon$. Thus, the composition of
gravastar can be specified by three geometries based on the
following EoS
\begin{itemize}
\item Internal geometry $(\mathcal{G}_{1})$ $\Rightarrow$ $p+\rho=0$
for $0\leq r<\mathcal{R}_{1}$,
\item Thin-shell $(\mathcal{G}_{2})$ $\Rightarrow$ $p=\rho$ for
$\mathcal{R}_{1}\leq r\leq \mathcal{R}_{2}$,
\item External geometry $(\mathcal{G}_{3})$ $\Rightarrow$ $p=\rho=0$
for $\mathcal{R}_{2}<r$.
\end{itemize}

\subsubsection{The Inner Geometry}
\begin{figure}\center
\epsfig{file=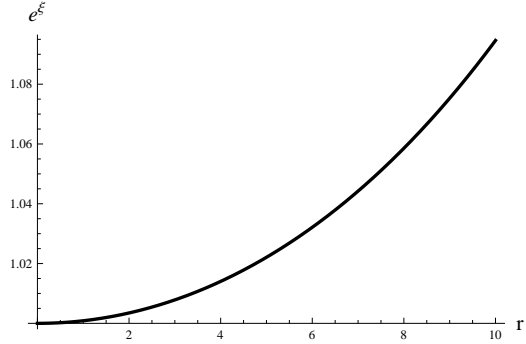,width=0.5\linewidth} \caption{Plot of
$e^{\xi(r)}$ against $r (km)$ with $\alpha=0.2$ and
$\rho_{c}=0.0001$.}
\end{figure}

In this domain, following the concept of Mazur and Mottola \cite{1},
we take the EoS $\mathcal{W}\rho=p$ with $\mathcal{W}$ being the
parameter of EoS and on setting $\mathcal{W}=-1$, we obtain
$-\rho=p$ which manifests the dark energy EoS. The negative pressure
in this EoS acts radially outward from the core of spherical object
to overcome the inward directed pull by the thin-shell. Hence, the
dark energy is responsible for this repulsive force in the inner
domain of gravastar. With the help of this EoS, Eq.(\ref{18}) yields
$\rho=\rho_{c}(\mathrm{constant})$, and hence
\begin{equation}\label{19}
p=-\rho=-\rho_{c}.
\end{equation}
This equation demonstrates that throughout the interior geometry,
the pressure and matter density remain constant. Inserting this
equation in Eq.(\ref{15}), the other metric function $e^{\xi(r)}$ is
obtained as
\begin{equation}\label{20}
e^{-\xi(r)}=1-\frac{4(\alpha+2\pi)r^{2}\rho_{c}}{3}+\frac{C_{1}}{r},
\end{equation}
where $C_{1}$ is the integration constant and we put $C_{1}=0$ to
obtain the regular solution at the center, i.e., $r=0$.
Consequently, Eq.(\ref{20}) becomes
\begin{equation}\label{21}
e^{-\xi(r)}=1-\frac{4(\alpha+2\pi)r^{2}\rho_{c}}{3}.
\end{equation}
The analysis of $e^{\xi(r)}$ against radial component is presented
in Figure \textbf{1} for particular values of $\alpha$ and
$\rho_{c}$ which interprets that this metric function is singularity
free as well as regular inside the gravastar. In the interior
geometry, the active gravitational mass can be evaluated through the
following equation
\begin{equation}\label{22}
\mathcal{M}(\mathcal{R})=4\pi\int_{0}^{\mathcal{R}_{1}=\mathcal{R}}
r^{2}\big[\rho-\frac{\alpha}{8\pi}(p-3\rho)\big]dr=\frac{2(2\pi+\alpha)
\mathcal{R}^{3}\rho_{c}}{3}.
\end{equation}
The behavior of gravitational mass is displayed in Figure \textbf{2}
which shows that the metric is free from any singularity at the
center of gravastar.
\begin{figure}\center
\epsfig{file=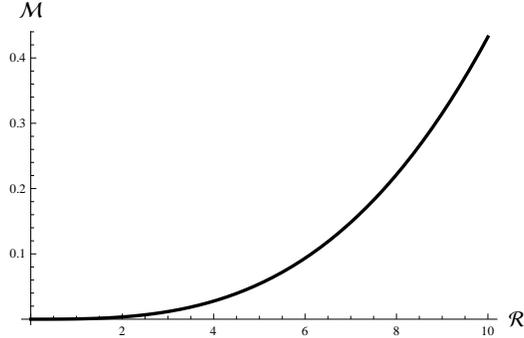,width=0.5\linewidth} \caption{Evolution of
gravitational mass ($M_{\odot}$) versus $\mathcal{R} (km)$ with
$\alpha=0.2$ and $\rho_{c}=0.0001$.}
\end{figure}

\subsubsection{The Intermediate Shell}

Here, we assume that the intermediate gravastar thin-shell consists
of stiff/ ultra-relativistic matter obeying the EoS $p=\rho$. In
connection with the cold baryonic universe, Zel'dovich \cite{38}
introduced the idea of such sort of matter distribution specified as
stiff fluid configuration. In this scenario, it may arise due to the
gravitational quanta at negligible temperature or because of some
thermal perturbations with null chemical potential \cite{1}. This
fluid has been utilized by several researchers \cite{39}-\cite{43}
to resolve different astrophysical as well as cosmological problems.
Implementation of stiff fluid EoS in Eqs.(\ref{15}) and (\ref{16})
leads to
\begin{equation}\label{23}
e^{-\xi(r)}=\frac{e^{Ar^{2}}-AC_{2}}{Ar^{2}e^{Ar^{2}}},
\end{equation}
where the integration constant is presented by $C_{2}$ which may be
evaluated by utilizing the boundary condition. Substituting
Eq.(\ref{23}) along with the stiff fluid EoS and Kuchowicz metric
function in Eq.(\ref{17}), the matter density and pressure turn out
to be
\begin{equation}\label{24}
p=\rho=-\frac{1-AC_{2}e^{-Ar^{2}}-Ar^{2}-A^{2}r^{4}}{2(4\pi+\alpha)Ar^{4}}.
\end{equation}

\subsubsection{The Exterior Geometry and Boundary Conditions}
\begin{figure}\center
\epsfig{file=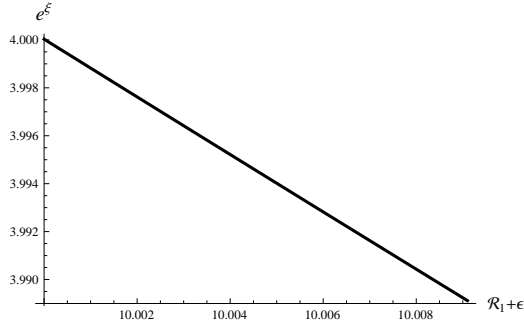,width=0.5\linewidth}\caption{Variation of
$e^{\xi(r)}$ versus thickness of the thin-shell.}
\end{figure}

It is well-known that the Einstein field equations are highly
nonlinear in nature but possess one most significant solution
outside the spherical object of total mass $M$, known as the
Schwarzschild solution. The exterior geometry of gravastar
satisfying the $\rho=p=0$ EoS can be characterized by the
Schwarzschild metric given by
\begin{equation}\label{25}
ds^{2}_{+}=\left(1-\frac{2M}{r}\right)dt^{2}-\frac{dr^{2}}{\left(1
-\frac{2M}{r}\right)}-r^{2}d\theta^{2}-r^{2}\sin^{2}\theta
d\phi^{2}.
\end{equation}
The structure of gravastar has two boundaries, one interrelates the
internal geometry and intermediate thin-shell while the other
connects the thin-shell and the external metric. For a viable
system, the metric potentials should be continuous at these
boundaries. In order to obtain the values of unknown constants $A$,
$B$ and $C_{2}$, we match the metric potentials at the interfaces as
follows
\begin{eqnarray}\label{26}
A&=&\frac{-M}{\mathcal{R}_{2}^{2}(2M-\mathcal{R}_{2})},\\\label{27}
B&=&\sqrt{\Big(1-\frac{2M}{\mathcal{R}_{2}}\Big)e^{\frac{M}{2M
-\mathcal{R}_{2}}}},\\\label{28}
C_{2}&=&\frac{e^{A\mathcal{R}_{2}^{2}}}{A}-\mathcal{R}_{2}^{2}
e^{A\mathcal{R}_{2}^{2}}\Big(1-\frac{2M}{\mathcal{R}_{2}}\Big).
\end{eqnarray}
To determine the values of these constants, we consider mass of the
gravastar $M=3.75M_{\odot}$, inner radius $\mathcal{R}_{1}= 10 km$
and outer radius $\mathcal{R}_{2}=10.009 km$ \cite{44} which provide
$A=0.01491932683 km^{-2}$, $B=0.2371387429$ and $C_{2}=186.8400519$.
The evolution of the metric potential (\ref{23}) as well as the
matter density and pressure (\ref{24}) by inserting the above
calculated values of unknowns are displayed in Figures \textbf{3}
and \textbf{4}, respectively. We see that the metric function does
not possess any singularity and pressure or energy density exhibits
positive values throughout the thin-shell of gravastar which sharply
decrease with thickness of the shell.
\begin{figure}\center
\epsfig{file=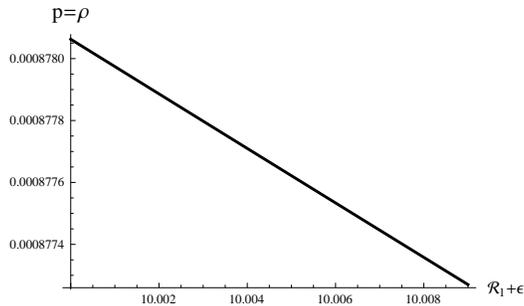,width=0.5\linewidth} \caption{Behavior of
$p=\rho$ $(km^{-2})$ versus thickness with $\alpha=0.2$.}
\end{figure}

\subsubsection{Darmois-Israel Matching Constraint}

The study of astrophysical objects demands that there must be a
smooth connection between inner and outer geometries. We have
already discussed that the structure of gravastar depends upon three
domains, i.e., $\mathcal{G}_{1}$, $\mathcal{G}_{2}$ and
$\mathcal{G}_{3}$ in which $\mathcal{G}_{2}$ acts as a junction
between $\mathcal{G}_{1}$ and $\mathcal{G}_{3}$. Implementing the
Darmois-Israel formalism \cite{45,46}, there must be a smooth
relation between $\mathcal{G}_{1}$ and $\mathcal{G}_{3}$ geometries
of the gravastar. At the junction ($r=\mathcal{R}$), though the
coefficients of the metric are continuous but their differentials
may not be continuous at this interface. In order to evaluate the
stress-energy of the surface ($\mathcal{S}_{ab}$), we consider the
Lanczos equation \cite{47,48} expressed by
\begin{equation}\label{29}
\mathcal{S}^{a}_{b}=\frac{1}{8\pi}\left(\alpha^{a}_{b}
\mu^{\lambda}_{\lambda}-\mu^{a}_{b}\right),
\end{equation}
where $\mu_{ab}=\mathcal{K}^{+}_{ab}-\mathcal{K}^{-}_{ab}$ yields
the discontinuous form of the extrinsic curvature. Here, the $-$ and
$+$ signs manifest the inner and outer domains of gravastar,
respectively.

The extrinsic curvature along with two regions of the intermediate
shell is characterized by
\begin{equation}\label{30}
\mathcal{K}^{\pm}_{ab}=-\Upsilon^{\pm}_{\nu}\left[\frac{\partial^{2}
x^{\nu}}{\partial\zeta^{a}\partial\zeta^{b}}+\Gamma^{\nu}_{\alpha\beta}
\frac{\partial x^{\alpha}\partial x^{\beta}}
{\partial\zeta^{a}\partial\zeta^{b}}\right],
\end{equation}
where $\zeta^{a}$ symbolizes the intrinsic coordinates on the
thin-shell while $\Upsilon^{\pm}_{\nu}$ reveal the unit normals at
hypersurface. For the line element
\begin{equation}\label{31}
ds^{2}=\mathcal{J}(r)dt^{2}-\frac{dr^{2}}{\mathcal{J}(r)}-r^{2}d\theta^{2}
-r^{2}\sin^{2}\theta d\phi^{2},
\end{equation}
the unit normals are described as
\begin{equation}\label{32}
\Upsilon^{\pm}_{\nu}=\pm\left|g^{\alpha\beta}\frac{\partial\mathcal{J}
}{\partial x^{\alpha}}\frac{\partial\mathcal{J}}{\partial
x^{\beta}}\right|^{-\frac{1} {2}}\frac{\partial\mathcal{J}}{\partial
x^{\nu}},\quad \textrm{satisfying}
\quad\Upsilon^{\nu}\Upsilon_{\nu}=1.
\end{equation}
Adopting the Lanczos equations, the mathematical expression of
$\mathcal{S}_{ab}=$diag$(\Omega,$
$-\mathcal{P},-\mathcal{P},-\mathcal{P})$ is acquired with $\Omega$
being the surface energy density and $\mathcal{P}$ being the surface
pressure. The forms of $\Omega$ and $\mathcal{P}$ are demonstrated
by
\begin{equation}\label{33}
\Omega=-\frac{1}{4\pi\mathcal{R}}
\left[\sqrt{\mathcal{J}}\right]^{+}_{-},\quad
\mathcal{P}=\frac{-\Omega}{2}+\left[\frac{\mathcal{J}'}
{16\pi\sqrt{\mathcal{J}}}\right]^{+}_{-}.
\end{equation}
Inserting $\mathcal{G}_{1}$ and $\mathcal{G}_{3}$ geometries of
gravastar in the above forms, we obtain
\begin{eqnarray}\label{34}
\Omega&=&\frac{1}{4\pi\mathcal{R}}\Big[\sqrt{1-\frac{4(2\pi+\alpha)
\mathcal{R}^{2}\rho_{c}}{3}}-\sqrt{1-\frac{2M}
{\mathcal{R}}}\Big],\\\label{35}
\mathcal{P}&=&-\frac{1}{8\pi\mathcal{R}}\Bigg[\frac{1-\frac{8(2\pi+\alpha)
\mathcal{R}^{2}\rho_{c}}{3}}{\sqrt{1-\frac{4(2\pi+\alpha)
\mathcal{R}^{2}\rho_{c}}{3}}}-\frac{1-\frac{M}
{\mathcal{R}}}{\sqrt{1-\frac{2M}{\mathcal{R}}}}\Bigg].
\end{eqnarray}
The graphical description of $\Omega$ and $\mathcal{P}$ is exhibited
in Figure \textbf{5}. It is found that both the parameters display
finite as well as positive behavior within the shell which assures
the creation and stability of the intermediate thin-shell of
gravastar. Employing Eq.(\ref{34}), we derive the surface mass for
the thin-shell of gravastar as
\begin{equation}\label{36}
\mathcal{M}_{shell}=4\pi\mathcal{R}^{2}\Omega=\mathcal{R}
\Big[\sqrt{1-\frac{4(2\pi+\alpha)\mathcal{R}^{2}\rho_{c}}{3}}
-\sqrt{1-\frac{2M}{\mathcal{R}}}\Big],
\end{equation}
which provides the total mass in the form of mass of gravastar
thin-shell as follows
\begin{equation}\label{37}
M=\frac{2(2\pi+\alpha)\mathcal{R}^{3}\rho_{c}}{3}
-\frac{\mathcal{M}_{shell}^{2}}{2\mathcal{R}}+\mathcal{M}_{shell}
\sqrt{1-\frac{4(2\pi+\alpha)\mathcal{R}^{2}\rho_{c}}{3}}.
\end{equation}
\begin{figure}\center
\epsfig{file=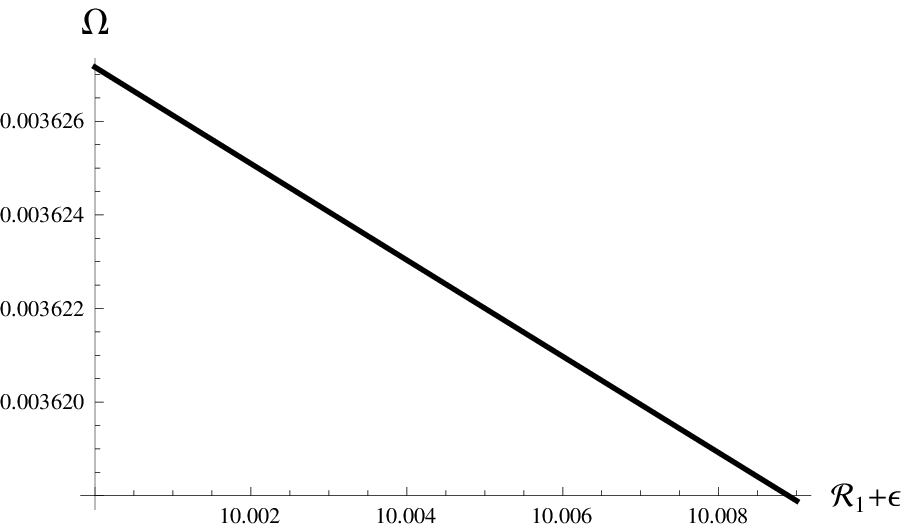,width=0.45\linewidth}
\epsfig{file=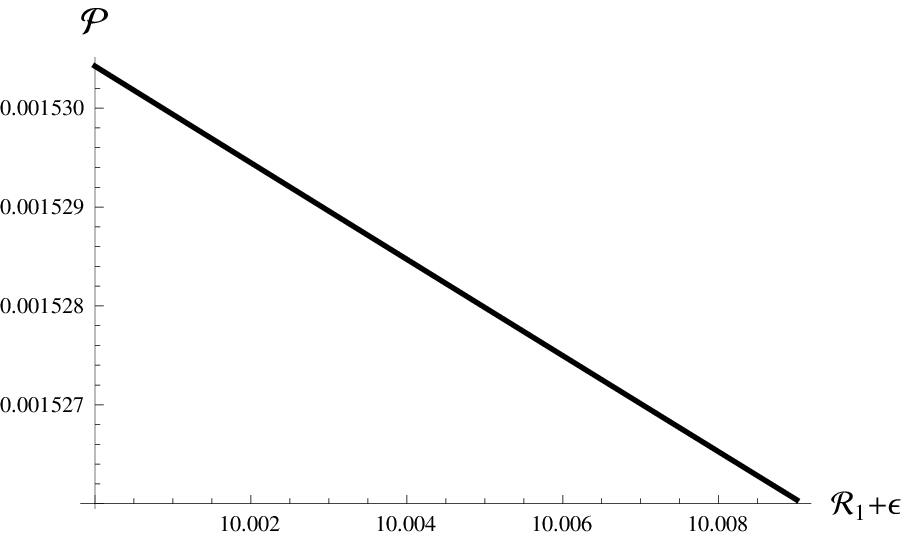,width=0.45\linewidth} \caption{Variation of
$\Omega$ and $\mathcal{P}$ versus thickness of thin-shell with
$\alpha=0.2$ and $\rho_{c}=0.0001$.}
\end{figure}

\section{Some Features of Gravastar Shell}

This section inspects some necessary features of gravastar like EoS
parameter, proper length, energy-thickness relationship and surface
redshift that completely portray the intermediate thin-shell of
gravastar.

\subsection{The EoS Parameter}

For the intermediate geometry of gravastar, Eqs.(\ref{34}) and
(\ref{35}) lead to the parameter of EoS at $r=\mathcal{R}$ as
\begin{equation}\label{38}
\mathcal{W}(\mathcal{R})=\frac{\mathcal{P}}{\Omega}=
\frac{\Bigg[\frac{1-\frac{8(2\pi+\alpha)\mathcal{R}^{2}
\rho_{c}}{3}}{\sqrt{1-\frac{4(2\pi+\alpha)\mathcal{R}^{2}
\rho_{c}}{3}}}-\frac{1-\frac{M}{\mathcal{R}}}{\sqrt{1
-\frac{2M}{\mathcal{R}}}}\Bigg]}{2\Big[\sqrt{1-\frac{2M}
{\mathcal{R}}}-\sqrt{1-\frac{4(2\pi+\alpha)
\mathcal{R}^{2}\rho_{c}}{3}}\Big]}.
\end{equation}
The positive matter density and pressure always provide positive
value of $\mathcal{W}$. For EoS parameter to be real, the required
constraints are $\frac{2M}{\mathcal{R}}<1$ and $\frac{4(2\pi+\alpha)
\mathcal{R}^{2}\rho_{c}}{3}<1$. Further, by expanding the
square-root terms both in the denominator as well as numerator of
Eq.(\ref{38}) in a binomial series along with the constraints
$\frac{M}{\mathcal{R}}\ll1$ and $\frac{4(2\pi+\alpha)
\mathcal{R}^{2}\rho_{c}}{3}\ll1$ and considering the terms up to
first order, we can attain
\begin{equation}\label{39}
\mathcal{W}(\mathcal{R})\approx\frac{3}{2
\Big[\frac{3M}{2(\alpha+2\pi)\mathcal{R}^{3}\rho_{c}}-1\Big]}.
\end{equation}
From this expression, the two possibilities for
$\mathcal{W}(\mathcal{R})$ can clearly be examined, i.e.,
$\mathcal{W}(\mathcal{R})$ is either negative if
$\frac{2(\alpha+2\pi)\rho_{c}}{3}>\frac{M}{\mathcal{R}^{3}}$, or
$\mathcal{W}(\mathcal{R})$ is positive if
$\frac{2(\alpha+2\pi)\rho_{c}}{3}<\frac{M}{\mathcal{R}^{3}}$.

\subsection{Proper Length for Intermediate Shell}
\begin{figure}\center
\epsfig{file=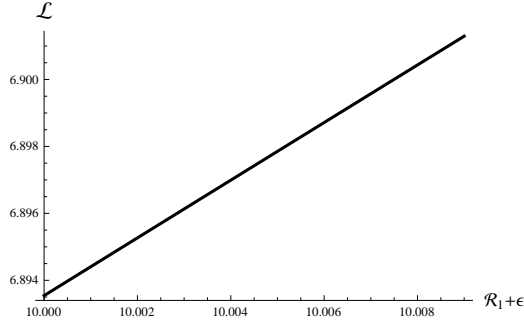,width=0.5\linewidth} \caption{Variation of
$\mathcal{L}$ $(km)$ versus thickness of thin-shell.}
\end{figure}

It has been assumed that the domain of stiff fluid gravastar shell
is situated at the boundary $r=\mathcal{R}$ which is specified
through the geometry $\mathcal{G}_{1}$. As, the thin-shell is
considered to be comprised of very small thickness, i.e.,
$\epsilon\ll1$, therefore, the geometry $\mathcal{G}_{3}$ originates
from the surface $r=\mathcal{R}+\epsilon$. Hence the proper length
($\mathcal{L}$) of the thin-shell is determined by the expression
\begin{equation}\label{40}
\mathcal{L}=\int^{\mathcal{R}+\epsilon}_{\mathcal{R}}\sqrt{e^{\xi(r)}}dr
=\Bigg[\frac{\ln\big\{e^{\frac{Ar^{2}}{2}}+\sqrt{e^{Ar^{2}}-AC_{2}}\big\}}
{\sqrt{Ae^{\frac{Ar^{2}}{2}}}}\Bigg]^{\mathcal{R}+\epsilon}_{\mathcal{R}}.
\end{equation}
The graphical behavior of the proper length with respect to the
thickness of thin-shell gravastar is displayed in Figure \textbf{6}
which manifests the linear as well as increasing relation between
them.

\subsection{Energy within the Thin-Shell}

In the geometry $\mathcal{G}_{1}$, we have adopted the dark energy
EoS which expresses the negative matter density yielding the
presence of repulsive effects in the inner geometry of gravastar. In
$f(R,T)$ framework, the expression of energy in the intermediate
shell of gravastar turns out to be
\begin{eqnarray}\nonumber
\mathcal{E}&=&4\pi\int^{\mathcal{R}+\epsilon}_{\mathcal{R}} \rho
r^{2}dr=\frac{-4\pi}{4\pi+\alpha}\int^{\mathcal{R}
+\epsilon}_{\mathcal{R}}\Big[\frac{1-Ar^{2}-A^{2}r^{4}-AC_{2}
e^{-Ar^{2}}}{2Ar^{2}}\Big]dr,
\end{eqnarray}
whose integration leads to
\begin{eqnarray}\label{41}
\mathcal{E}=\frac{-4\pi}{4\pi+\alpha}\Bigg[\frac{C_{2}
e^{-Ar^{2}}}{r}-\frac{1}{Ar}-\frac{2Ar^{3}}{3}-r+\sqrt{AC_{2}\pi}
Erf[\sqrt{A} r]\Bigg]_{\mathcal{R}}^{\mathcal{R}+\epsilon},
\end{eqnarray}
where $Erf$ stands for the error function. The graphical analysis of
energy versus thickness demonstrates the attractive nature of energy
within the thin-shell and is shown in Figure \textbf{7}.
\begin{figure}\center
\epsfig{file=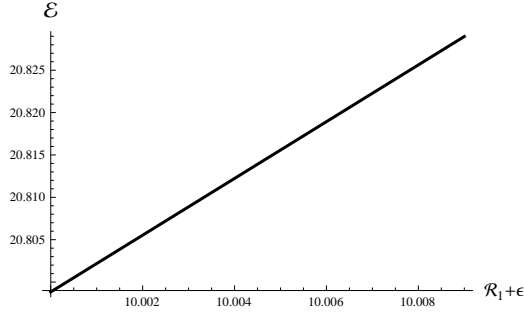,width=0.5\linewidth} \caption{Evolution of energy
$(km)$ versus thickness $(km)$ with $\alpha=0.2$.}
\end{figure}

\subsection{Surface Redshift within the Thin-Shell}

In the study of structure of gravastars, the surface redshift can be
regarded as one of the most significant sources of information
corresponding to the detection as well as stability of gravastars.
For static isotropic matter distribution, it is claimed that the
value of redshift parameter should not exceed 2 \cite{49,50}. The
expression of surface redshift is presented by \cite{34}
\begin{eqnarray}\label{42}
\mathcal{Z}_{s}=|g_{tt}|^{\frac{-1}{2}}-1=
\frac{1}{e^{\frac{Ar^{2}}{2}}B}-1.
\end{eqnarray}
The behavior of surface redshift (by substituting the numerical
values of $A$ and $B$) is exhibited in Figure \textbf{8} which
indicates that the value of redshift parameter lies within 1 inside
the intermediate thin-shell of gravastar. Hence, we can claim the
stable as well as physically consistent structure of gravastar.
\begin{figure}\center
\epsfig{file=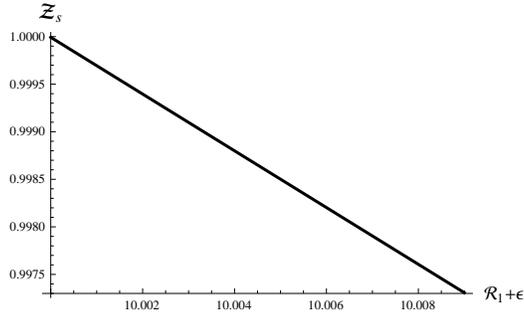,width=0.5\linewidth} \caption{Behavior of
$\mathcal{Z}_{s}$ against thickness of thin-shell.}
\end{figure}

\section{Concluding Remarks}

In this work, we have evaluated a new solution to study a novel
massive compact object named as gravastar in $f(R,T)$ framework by
considering the Kuchowicz metric function. This massive stellar
model can be regarded as a feasible substitute for a black hole. The
motivation behind the consideration of Kuchowicz metric potential is
that this metric function is purely non-singular and exhibits a
regular behavior within the geometry of gravastar. We have discussed
various properties of three geometries of gravastar with respect to
$R+2\alpha T$ functional form. We have obtained regular solutions
for the interior geometry at the center of gravastar and analyzed
different characteristics of the thin-shell for a particular value
of $\alpha$ which leads to physical viability of the gravastar.

In the interior geometry of gravastar, we have acquired solution of
the metric potential, matter density and pressure. Using the dark
energy EoS in the conservation equation, we have determined that the
matter density remains constant throughout the inner region. This
constant value of density is responsible for the outward pull to
keep the gravastar in stable state. The evolution of the metric
function $e^{\xi(r)}$ (Figure \textbf{1}) remains finite at the core
of interior geometry. The behavior of gravitational mass has also
been analyzed which becomes zero at the origin and positive within
the interior domain. Thus, the considered metric avoids any type of
central singularity.

In the intermediate thin-shell of gravastar, we have assumed the
ultra-relativistic matter configuration and solved the field
equations to evaluate the metric function. For the exterior
geometry, we have employed the Schwarzschild metric and calculated
the values of unknown constants through boundary conditions. We have
also determined the surface density and pressure by utilizing the
Darmois-Israel matching constraints. We have then inspected some
important aspects of gravastar like variation of matter density or
pressure, EoS parameter, proper length, energy and surface redshift
associated with the thickness of the thin-shell. All these
parameters manifest the stable structure of gravastar and
demonstrate the physical acceptability of our considered metric
function as well as $f(R,T)$ model.

The exact solutions of the field equations play a crucial role to
discuss the composition of astrophysical objects. Das et al.
\cite{29} presented different features of gravastar in $f(R,T)$
framework and evaluated a set of complete solutions for three
different regions with their corresponding EoS. Shamir and Ahmad
\cite{30} solved the $f(\mathcal{G},T)$ field equations to obtain
non-singular solutions of gravastar. Yousaf et al. \cite{51}
observed the effects of electromagnetic field on the formation of
gravastar for $R+2\chi T$ model of $f(R,T)$ gravity. Yousaf and his
collaborators \cite{52} discussed the significance and
characteristics of gravastar under some specific constraints in
$f(R,T,R_{\mu\nu}T^{\mu\nu})$ gravity. Bhatti et al. \cite{53}
explored the non-singular spherical model with a particular EoS in
$f(R,G)$ gravity. In all these works, the authors have not used any
particular metric potential to inspect physically viable structure
of gravastar in modified theories.

In the context of GR, Ghosh et al. \cite{34} investigated the
regions of gravastar by adopting the Kuchowicz metric potential.
They analyzed various features of gravastar graphically and found
stable structure under the effect of their considered non-singular
metric potential. We have obtained new exact solutions of $f(R,T)$
field equations by implementing Kuchowicz metric potential to
describe the geometries of gravastar. We conclude that the
construction of gravastar like stellar model admitting the Kuchowich
metric potential in this framework seems to be physically consistent
as well as viable similar to GR \cite{34}. Hence, $f(R,T)$ gravity
can provide stable as well as acceptable structures of massive
compact objects under the influence of specific metric function same
as in GR. It would be interesting to discuss different regions of
gravastar with respect to several other types of metric functions in
$f(R,T)$ theory as well as in other modified gravitational theories.

\end{document}